\documentclass[conference]{IEEEtran} 
\IEEEoverridecommandlockouts
\usepackage{amsmath,graphicx,times,booktabs, tabularx}
\usepackage[hyphens]{url}
\usepackage{multirow}
\usepackage{color,soul}
\usepackage{flushend}

\begin{document}
\title{Multichannel Sound Event Detection \\Using 3D Convolutional Neural Networks \\for Learning Inter-channel Features}

\author{\IEEEauthorblockN{Sharath Adavanne\textsuperscript{1}, Archontis Politis\textsuperscript{2}, Tuomas Virtanen\textsuperscript{1} \thanks{The research leading to these results has received funding from the European Research Council under the European Union’s H2020 Framework Programme through ERC Grant Agreement 637422 EVERYSOUND. The authors also wish to acknowledge CSC-IT Center for Science, Finland, for computational resources}}
\IEEEauthorblockA{\textsuperscript{1}Laboratory of Signal Processing, Tampere University of Technology, Finland\\
Email: firstname.lastname@tut.fi \\
\textsuperscript{2}Department of Signal Processing and Acoustics, Aalto University, Finland\\
Email: archontis.politis@aalto.fi}
}

\maketitle
\begin{abstract}
In this paper, we propose a stacked convolutional and recurrent neural network (CRNN) with a 3D convolutional neural network (CNN) in the first layer for the multichannel sound event detection (SED) task. The 3D CNN enables the network to simultaneously learn the inter- and intra-channel features from the input multichannel audio. In order to evaluate the proposed method, multichannel audio datasets with different number of overlapping sound sources are synthesized. Each of this dataset has a four-channel first-order Ambisonic, binaural, and single-channel versions, on which the performance of SED using the proposed method are compared to study the potential of SED using multichannel audio. A similar study is also done with the binaural and single-channel versions of the real-life recording TUT-SED 2017 development dataset. The proposed method learns to recognize overlapping sound events from multichannel features faster and performs better SED with a fewer number of training epochs. The results show that on using multichannel Ambisonic audio in place of single-channel audio we improve the overall F-score by 7.5~\%, overall error rate by 10~\% and recognize 15.6~\% more sound events in time frames with four overlapping sound sources.

% In this paper, we assess the performance of using multichannel audio in place of single-channel audio for polyphonic sound event detection (SED). Multichannel audio datasets with different number of overlapping sound sources are synthesized, where each dataset has a four-channel first-order Ambisonic, binaural, and single-channel versions. The proposed stacked convolutional and recurrent neural network with a 3D convolutional layer is trained individually on the synthesized multichannel datasets. This 3D convolution enables the network to simultaneously learn the inter- and intra-channel features from multichannel input data. The SED performances with single-channel, binaural and Ambisonic audio are further compared to study the potential of SED using multichannel audio. A similar study is also done with the binaural and single-channel versions of the real-life recording TUT-SED 2017 development dataset. The results show that on using multichannel Ambisonic audio in place of single-channel audio we improve the overall F-score by 7.5~\%, overall error rate by 10~\% and recognize 15.6~\% more sound events in time frames with four overlapping sound sources.
\end{abstract}

\IEEEpeerreviewmaketitle

\section{Introduction}
\label{sec:intro}

Sound event detection (SED) is the task of recognizing the sound events and their respective temporal start and end time in an audio recording. Sound events in real life do not always occur in isolation but tend to considerably overlap with each other. Recognizing such overlapping sound events is referred as polyphonic SED. Applications of such polyphonic SED are numerous. Recognizing sound events like alarm and glass breaking can be used for surveillance~\cite{surveillance_audio,surveillance}. Automatic detection of road accidents can ensure quick intervention of emergency teams~\cite{Foggia_TITS2015}. Environmental sound event detection can be used for monitoring biodiversity~\cite{environmentalSED,Marques2012,Furnas2014}. Further, SED can be used for automatically annotating audio datasets, and the sound events recognized can be used as a query for similar content retrieval.
% Valenzise_AVSS2007

Polyphonic SED using single-channel audio has been studied extensively. Different approaches have been proposed using supervised classifiers like Gaussian mixture model - hidden Markov model~\cite{Mesaros2010_EUSIPCO}, fully-connected networks~\cite{emre2015}, convolutional neural networks (CNN)~\cite{Zhang2015,Phan2016}, and recurrent neural networks (RNN)~\cite{Adavanne_DCASE2016,Hayashi_TASLP2017,Zohrer_INTERSPEECH2017}. More recently, the state of the art method for polyphonic SED was proposed in~\cite{emre_TASLP2016}, where the log mel-band energy feature was used with a convolutional recurrent neural network (CRNN) architecture.

Recognizing overlapping sound events using a single-channel audio is a challenging task. These overlapping sound events can potentially be recognized better with multichannel audio. One of the first methods to use multichannel audio for SED was proposed in~\cite{temko2007}, which performed SED on each of the audio channels separately and the combined likelihoods across channels were used for the final prediction. More recently the state of the art CRNN network for single-channel SED~\cite{emre_TASLP2016} was extended for multichannel features and multiple feature classes in~\cite{Adavanne2017}. It was shown that the performance of SED improves on using the binaural audio instead of the single-channel audio version of the same dataset. In this regard,~\cite{Adavanne2017} also proposed binaural audio features exploiting the inter-aural intensity and time differences. In the network proposed by~\cite{Adavanne2017} the CNNs were used as feature extractors that learned just the intra-channel information from the input multichannel audio features, while the RNNs which followed the CNNs were learning the inter-channel information. In this paper, we propose to learn both the inter- and intra-channel information within the CNN layer. We implement this by using a 3D CNN~\cite{Tran_ICCV2015} as the first layer of the network. This enables the method to learn both inter- and intra-channel information from the input multichannel audio within the CNN layers for no additional parameters in comparison to~\cite{Adavanne2017}.

The hardware devices for smart homes, virtual reality content creation, modern hearing aids and surveillance sensors have more than one microphone in them. By using all the multichannel audio available from these devices we can potentially improve the polyphonic SED, and this improvement can additionally enhance the overall performance of these devices. Although~\cite{Adavanne2017} showed that using binaural audio in place of single-channel audio improves the performance of SED, there is no other conclusive work that studies the potential of SED with more than two-channel of audio. Besides, in order to carry out such a study, there are no publicly available data. Moreover, collecting and annotating such a dataset for SED is a tedious, expensive and time-consuming task. In order to assess the necessity of collecting such a dataset, in this paper, we synthesize three multichannel audio datasets with up to one, up to three and up to six temporally overlapping sound sources. The multichannel audio in each of the datasets is a four-channel first-order Ambisonic (FOA) audio. Additionally, we perform binauralization with real head related transfer function (HRTF) to obtain binaural version from the FOA audio, and further used the omnidirectional channel of FOA as the single-channel version. Experiments are carried out on these datasets to understand the extent of improvement we can achieve by using multichannel audio over the current state of the art SED methods using single-channel and binaural audio. Based on the results obtained we can decide to invest in the collection of real-life multichannel dataset. Furthermore, in order to compare the consistency of results obtained with the synthetic dataset, we perform similar experiments on the real-life recordings TUT-SED 2017 dataset~\cite{dcase2017}, that consists of only the single-channel and binaural audio.    

%In particular, for the polyphonic SED task, we synthesize multichannel datasets with different number of overlapping sound sources. We individually train the multichannel CRNN with 3D CNN as the first layer with a) the multichannel, binaural, and the single-channel version of the synthesized dataset, and b) binaural and the single-channel version of the real-life dataset. The performance of SED using single, binaural and multichannel audio is studied to assess the necessity of real-life multichannel audio dataset for polyphonic SED.

The paper is organized as follows: Section~\ref{sec:method} describes the feature extraction details and the proposed neural network. The datasets used, metric for evaluation, the baseline method and the evaluation procedure are explained in Section~\ref{sec:eval}. Finally, the results and discussion are presented in Section~\ref{sec:results}.

\begin{figure*}
  \centering
  \centerline{\includegraphics[width=\linewidth]{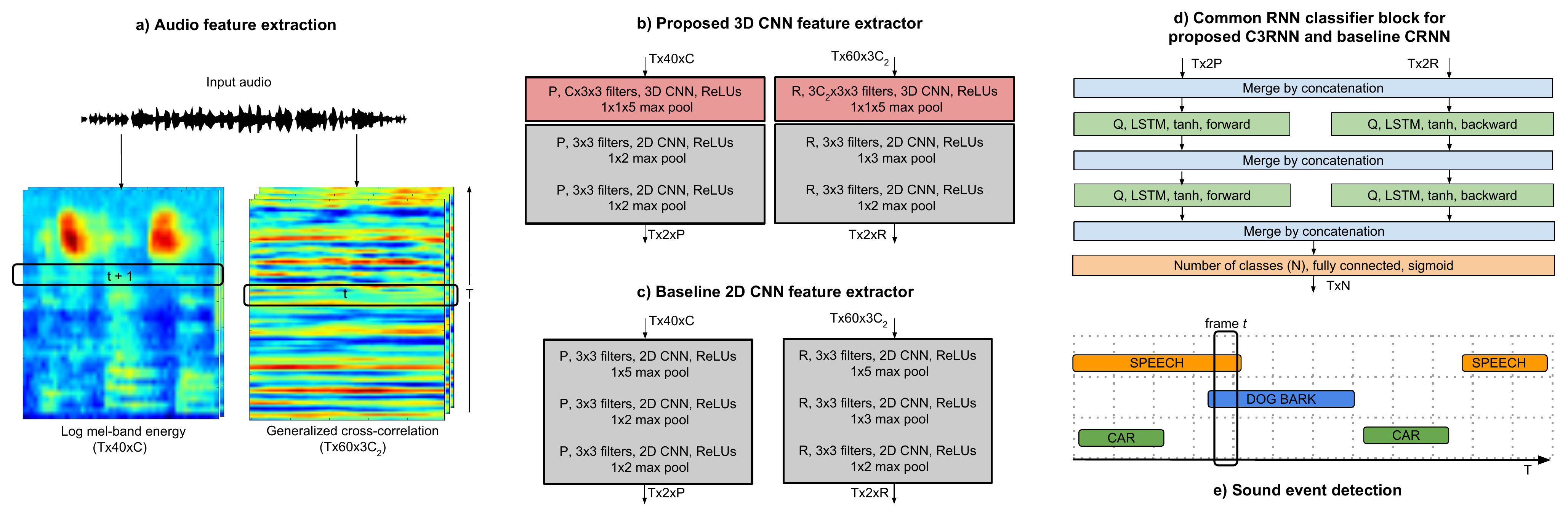}}
  \caption{The proposed C3RNN (a+b+d+e) and baseline CRNN (a+c+d+e) stacked convolutional and recurrent neural network architectures for multichannel polyphonic sound event detection.}
  \label{fig:crnn}
\end{figure*}

\section{Method}
\label{sec:method}
The proposed multichannel SED method is shown in Figure~\ref{fig:crnn}. The input to the method is either a single-channel or multichannel audio. For single-channel audio input, we use just the log mel-band energy feature. In the case of multichannel input, the log mel-band energy feature is extracted in each of the channels; additionally, generalized cross-correlation with phase transform (GCC-PHAT)~\cite{GCC_PHAT} feature is extracted between each channel pair of the multichannel audio. These audio features are fed to a multichannel neural network architecture that maps them to the activities of the sound event classes in the dataset. The output of the neural network is in the continuous range of $[0, 1]$ for each of the sound event classes and corresponds to the probability of the particular sound class being active in the frame. This continuous range output is further thresholded to obtain the final binary decision of the sound event class being active or absent in each frame. In general, the proposed method takes a sequence of frame-wise audio features as the input and predicts the activity of the target sound event classes for each of the input frames. The detailed description of feature extraction and the neural network is presented below.

\subsection{Feature extraction}
\label{ssec:feat}
\subsubsection{Log mel-band energy}
\label{sssec:mbe}
Previously, single-channel SED methods have been using log mel-band energies ($mbe$) and have shown to be effective for the task~\cite{emre2015,Adavanne_DCASE2016, emre_TASLP2016, Adavanne2017,Jeong_DCASE2017,Lee_DCASE2017}. Additionally, in the case of SED with binaural audio, $mbe$ extracted from the two channels was proposed in~\cite{Adavanne_DCASE2016}. This was motivated from the inter-aural intensity difference (IID) used by the human auditory system to localize and recognize overlapping sound events. A neural network that is capable of performing linear operations (which includes the difference operation) can obtain information similar to the IID from the binaural $mbe$. More recently, the binaural $mbe$ was shown to improve the performance of SED even on larger binaural datasets~\cite{Adavanne2017}. Motivated from this, we continue to use $mbe$ feature extracted from all the input channels in this paper.

In case of a single-channel audio input, we extract $mbe$ in 40 ms windows with 50\% overlap and refer to it as $mbe$-$mono$. We use 40 mel-bands in the frequency range of 0-22500 Hz. For multichannel audio we extract $mbe$ in each of the channels, and refer to it as $mbe$-$bin$ for binaural and $mbe$-$ambi$ for four-channel FOA audio. For a sequence length of $T$ frames, the $mbe$ feature has a general dimension of $T\times40\times C$, where $C$ is the number of channels, $C=1$ for $mbe$-$mono$, $C=2$ for $mbe$-$bin$ and $C=4$ for $mbe$-$ambi$.

\subsubsection{Generalized cross correlation with phase transform}
\label{sssec:mbe}
In the case of binaural audio,~\cite{Adavanne2017} proposed to represent similar information as the inter-aural time difference (ITD) in humans using the generalized cross correlation with phase transform ($gcc$). It was shown that the SED methods can benefit with $gcc$ for overlapping sound events. Motivated from this, we continue to use $gcc$ in this paper. Similar to~\cite{Adavanne2017}, we extract $gcc$ in three resolutions, 120, 240, and 480 ms as
\begin{equation}
 \label{eqn:1}
 R(\Delta_{12},t) = \sum_{k=0}^{K-1} \frac{  X_1(k,t) \cdot X_2^\ast(k,t)}{\arrowvert X_1(k,t)\arrowvert \arrowvert X_2(k,t)  \arrowvert} e^{ \frac{i 2\pi k \Delta_{12}}{N}},
\end{equation} 
where, $X_1$ and $X_2$ are the FFT coefficients of the two-channels between which the $gcc$ is calculated. $X_1(k,t)$ is the coefficient at time frame $t$ and $k$th frequency bin, of the total $K$ bins. $gcc$ per frame given by $R(\Delta_{12},t)$ is extracted for delays $\Delta_{12}$ in the range $[-\tau_\textrm{max}, \tau_\textrm{max}]$, where $\tau_\textrm{max}$ is the maximum sample delay for a sound wave to travel between the pair of microphones recording audio. In order to have a factorisable feature length for max pooling in the neural network, 60 $gcc$ values are chosen in the range $\Delta_{12} \in [-29, 30]$ lag for each of the three multi-resolution. For a sequence length of $T$ frames, the $gcc$ feature is of the general dimension $T\times60\times3\binom{C}{2}$, where $\binom{C}{2}$ is the number of possible pair-of-two combinations for the $C$ channels of audio (denoted in Figure~\ref{fig:crnn} as $C_2$) and $3$ is the number of resolutions in which $gcc$ was extracted. In the case of binaural audio ($gcc$-$bin$) this results in $T\times60\times3$ and for Ambisonic audio ($gcc$-$ambi$) this amounts to $T\times60\times18$. 

%  given by $\tau_\textrm{max} = Fs * d / c$, where $Fs$ is the sampling frequency, $d$ is the distance between the pair of microphones and $c$ is the speed of sound.

\subsection{Neural network}
\label{ssec:dnn}

The input to the proposed method is $T\times40\times C$ dimensional $mbe$ and $T\times60\times3\binom{C}{2}$ dimensional $gcc$ features as shown in Figure~\ref{fig:crnn}. Based on the task of single or multichannel SED, the network is fed with the respective feature sequence. 

Separate CNN branches are used to learn local shift-invariant features from each of the input features $mbe$ and $gcc$. The first CNN layer in each CNN branch consists of a 3D CNN, i.e., convolution over volumes. The receptive filters of 3D CNNs are of the size $D\times3\times3$ size, where $D=C$ for $mbe$ and $D=3\binom{C}{2}$ for $gcc$ feature. This joint learning of features along channel-time-frequency enables the network to learn both inter- and intra-channel features simultaneously within the first layer. The 3D CNN is followed by a sequence of 2D CNN layers with receptive filters of size $3\times3$. The output activation from both the CNN layers is padded with zeros to keep the dimension of the output the same as the input. Batch normalization~\cite{batchNorm} and max-pooling is performed after every layer of CNN along frequency axis to reduce the final dimension to $T\times2\times P$ for $mbe$ and $T\times2\times R$ for $gcc$, where $P$ and $R$ are the number of filters in the final layer of CNN in respective CNN branches. The CNN activations from the two branches are concatenated along feature axis and are fed to layers of bi-directional gated recurrent units (GRU), to learn long-term temporal activity patterns. This is followed by a layer of time-distributed fully-connected (dense) network. The final prediction layer has as many sigmoid units as the number of sound event labels in the dataset. We refer to this network as C3RNN in future.

The training is performed for 1000 epochs using Adam~\cite{adamKeras} optimizer, and binary cross-entropy loss between the reference sound class activities and the predicted ones. Dropout~\cite{Dropout} is used as a regularizer after every layer of the neural network to make it robust to unseen data. Early stopping is used to stop overfitting the network to training data. A threshold of 0.5 is used to obtain the binary decision from the sigmoid activations in the final prediction layer. Training is stopped if the error rate (see Section~\ref{ssec:metric}) on the test split does not improve for 100 epochs. The neural network implementation was done using the PyTorch~\cite{paszke2017automatic} library.

\section{Evaluation} 
\label{sec:eval}

\subsection{Dataset}
\label{ssec:data}
We evaluate the proposed C3RNN with four different datasets, one real-life audio TUT-SED 2017 Development dataset~\cite{dcase2017} and three synthetic datasets. The recordings of TUT-SED 2017 are binaural. In order to assess the performance of SED for more than two channels of audio we propose to use the synthetic datasets. 
\subsubsection{TUT-SED 2017 Development dataset}
This dataset was recorded in the street context using a binaural in-ear microphone at 24 bit and 44.1 kHz sampling rate. Each of the recordings is of the length 3-5 minutes, amounting to a total length of 70 minutes.  This dataset consists of manual annotations for sound event classes such as brakes squeaking, car, children, large vehicle, people speaking, and people walking. The dataset defines four-folds of training and testing splits for benchmarking. Further details of the dataset are given in~\cite{dcase2017}. Since the dataset has only two channels, we do not have $mbe$-$ambi$ and $gcc$-$ambi$ features for this dataset. The single-channel version is obtained by taking the mean of the binaural channels. 

\subsubsection{Synthetic dataset}
In order to assess the performance of SED in presence of more than two channels of audio, we generate synthetic datasets using the method proposed in~\cite{Adavanne_ICASSP2018}. Three separate anechoic multichannel datasets with a) no temporally overlapping sources ($O1$), b) maximum three overlapping sources ($O3$), and c) maximum six overlapping sources ($O6$) are synthesized. For each dataset, three sets of training and test split were generated, each with 500 and 100 recordings respectively. Every recording is of length 30 seconds and sampled at 44100 Hz. The dataset consists of only stationary point sources. Point sources are sound events which can be associated with a single spatial coordinate in the space, for example, a person speaking, or a phone ringing. Diffuse sources like ambient noise, wind breeze, etc. do not have a specific spatial coordinate and are therefore more difficult to synthesize spatially, hence we do not use them in this study. 

The audio recordings synthesized were of first-order Ambisonic (FOA) format. This is a commonly used format for spatial audio, especially in the virtual reality domain\footnote{https://developers.google.com/vr/concepts/spatial-audio}. The FOA consists of four channels of audio, commonly referred as W, X, Y, and Z channels, where, X, Y and Z channels represents the directive pressure-gradient recordings along the X, Y and Z axes of the Cartesian coordinate system respectively. The W channel corresponds to an omnidirectional microphone recording. In this paper, we use the W channel for our single-channel SED studies and all the four channels (W, X, Y, and Z) for our four-channel SED studies. We further perform binauralization of the four-channel audio using real head-related transfer functions (HRTF) to obtain the binaural version and used them for the binaural SED studies. The HRTFs were measured from one of the authors on a dense grid of directions under anechoic conditions, as detailed in~\cite{bolanos2012hrir}. For an overview on HRTF measurement techniques, and simulation of spatial sound scenes based on them, such as in this work, the reader is referred to~\cite{xie2013_hrtf}.

In order to synthesize these datasets, we use the isolated sound events from the DCASE 2016 task 2~\cite{dcase2016Task2}. This dataset consists of 11 sound event classes, with 20 examples each. The sound event classes include speech, cough, door slam, laughter, phone, knock. We chose 16 examples from each class randomly for training and four for the testing split. In order to synthesize a recording, each sound example was randomly associated with a spatial coordinate such that two temporally overlapping examples do not have the same spatial coordinate. Further, the magnitude of the sound examples was varied randomly to give the effect of varying distance from the microphone. Details of the synthesis procedure are given in~\cite{Adavanne_ICASSP2018}. 

\subsection{Metric}
\label{ssec:metric}
The proposed SED method is evaluated using the polyphonic SED metrics proposed in~\cite{metrics}. Particularly we use segment wise error rate (ER) and F-score calculated in one-second length segments. The F-score is calculated as 

\begin{equation}
F = \frac{2 \cdot \sum_{k=1}^{K} TP(k)}{2 \cdot \sum_{k=1}^{K}TP(k)+ \sum_{k=1}^{K}FP(k)+ \sum_{k=1}^{K}FN(k)},
\label{Eqn:F}
\end{equation}
where for each one-second segment $k$, $TP(k)$ is the number of true positives i.e., the number of sound event labels active in both predictions and ground truth. $FP(k)$ is the number of false positives i.e., the number of sound event labels active in predictions but inactive in ground truth. $FN(k)$ is the number of false negatives i.e., the number of sound event labels active in the ground truth but inactive in the predictions.

The error rate is measured as
\begin{align}
ER = \frac{\sum_{k=1}^{K} S(k) + \sum_{k=1}^{K} D(k) + \sum_{k=1}^{K} I(k)}{\sum_{k=1}^{K} N(k)},
\label{Eqn:ER}
\end{align}
where $N(k)$ is the total number of active sound events in the ground truth of segment $k$. The number of substitutions $S(k)$, deletions $D(k)$ and insertions $I(k)$ is measured using the following equations for each of the $K$ one second segments:
\begin{align}
S(k) = \min(FN(k), FP(k)) \\
D(k) = \max(0, FN(k)-FP(k)) \\
I(k) = \max(0, FP(k)-FN(k)) 
\end{align}

According to the Equations~(\ref{Eqn:F}) and~(\ref{Eqn:ER}), for an ideal SED method, ER is zero and F-score is one. In this paper, we report the F-score in percentage and hence the ideal F-score will be 100~\%.

\subsection{Baseline}
\label{ssec:baseline}
The proposed C3RNN is compared with the existing state of the art multichannel audio SED method proposed in~\cite{Adavanne2017}. Similar to the proposed C3RNN, the baseline method can perform SED with single-channel, binaural and multichannel audio. Previously, its performance has only been tested with single-channel and binaural audio. This method won~\cite{Adavanne_DCASE2017_binaural} the recently concluded IEEE Audio and Acoustic Signal Processing research challenge -- DCASE 2017 Task 3 for real life sound event detection~\cite{dcase2017}. In particular, it secured the first two positions among the 34 submitted methods. The first position was obtained with the $mbe$-$mono$ audio feature and a close second position with $mbe$-$bin$. This proves that the method is well suited for both single-channel and binaural SED baselines.

The baseline method shown in Figure~\ref{fig:crnn} is also based on a stacked convolutional and recurrent neural network (CRNN). In comparison to the proposed C3RNN, the method does not employ a 3D CNN, thus its CNN only learns intra-channel information, while the RNNs learn the inter-channel information. The rest of the inputs and the outputs of the baseline CRNN and proposed C3RNN are similar. In this paper, we consider  both the CRNN with single-channel and binaural audio features as the baselines, and further report the performance of CRNN with multichannel Ambisonic audio along with the C3RNN performance.

\subsection{Evaluation procedure}
\label{ssec:eval_proc}

In order to evaluate the performance of the proposed C3RNN with respect to the baseline CRNN on multichannel dataset, the two methods were trained individually using the single-channel, binaural, and Ambisonic audio features of the synthetic dataset and the single-channel, and binaural audio features of TUT-SED 2017 development dataset. We perform a hyper-parameter search on each of the dataset-feature combinations individually and assess the performance of SED using multichannel audio using the ER and F-scores on the test splits. The metric scores reported are the mean of five separate runs on the cross-validation splits.

In order to study the individual contribution of $gcc$ for the SED task, we performed an experiment of estimating the number of sound sources in every time frame using just the $gcc$ feature. The usage of $gcc$ for SED task was motivated from the idea that the relative time difference of arrival of two overlapping sound sources will be different, and this will be highlighted in the $gcc$ feature. In the proposed experiment of identifying the number of active sources, using just $gcc$ feature should have better accuracy than using only the $mbe$ feature. This would mean that the $mbe$ based SED methods will additionally benefit from using $gcc$.

We trained the proposed C3RNN with just $gcc$ feature as input and the number of active sound sources as the output. Similar training was done using just $mbe$ feature as input. When using a single feature in the proposed C3RNN method, for example, the $mbe$ feature, the CNN feature extractor branch for $gcc$ is removed, and only the CNN feature extractor branch for $mbe$ is used. Separate hyper-parameter search was done for the individual features randomly~\cite{Bergstra2012}, and the best configurations for both $gcc$-$ambi$ and $mbe$-$ambi$ features for synthetic dataset $O6$ had around 270 k trainable weights. Unlike the SED task which is a multi-label classification task (more than one sound event can be active in a given time frame), this experiment of estimating the number of sources is a multi-class classification task (mutually exclusive classes). Hence, for this experiment alone the output sigmoid activation was replaced with softmax, and the categorical cross entropy loss was used.

\section{Results and discussion} 
\label{sec:results}

A hyperparameter search was carried out with the proposed C3RNN and baseline CRNN for each combination of the dataset (synthetic $O1$, $O3$ and $O6$, and TUT-SED 2017 development dataset) and audio feature ($mbe$-$mono$, $mbe$-$bin$, $mbe$-$ambi$, $mbe$-$gcc$-$bin$, and $mbe$-$gcc$-$ambi$). In general, the hyperparameters remained the same for a given dataset, independent of the feature used. A sequence length of 128 frames, batch size of 32 and dropout of 0.35 gave the best results for all the feature and synthetic dataset combinations. For the TUT-SED 2017 dataset, the best results were obtained using a sequence length of 256 frames, a batch size of 128 and dropout of 0.2. A learning rate of $1\times10^{-4}$ gave the best results across datasets and audio features. The performance was not affected much by the exact number of CNN filters or GRU units. Across datasets, different number of CNN filters and GRU units were seen to give good evaluation metric scores. In the case of the synthetic $O1$ dataset, the optimal number of CNN filters for $mbe$ in each layer ($P$ in Figure~\ref{fig:crnn}) was 8, for $gcc$ in each layer ($R$ in Figure~\ref{fig:crnn}) it was 16 and the GRU units in each layer ($Q$ in Figure~\ref{fig:crnn}) was 8. Similarly $P=Q=16$ and $R=32$ for synthetic $O3$, $P=Q=32$ and $R=64$  for synthetic $O6$, and $P=Q=R=64$ for TUT-SED 2017 dataset. This correlation of increasing number of CNN filters and GRU units with increasing number of overlapping sound events in datasets shows that bigger neural networks are required for recognizing highly overlapped sound events.

\begin{table}[!t]
\centering
\caption{The evaluation metric scores for the sound event detection task using the proposed C3RNN and baseline CRNN with $mbe$ and $gcc$ audio feature for different overlapping sound events datasets.}
\label{T:error}
\begin{tabular}{lcc|cc|cc}
 & \multicolumn{2}{c}{$O1$} & \multicolumn{2}{c}{$O3$} & \multicolumn{2}{c}{$O6$} \\ \cline{2-7}
\multicolumn{1}{l|}{C3RNN} & ER & F & ER & F & ER & F \\ \hline
\multicolumn{1}{l|}{$mbe$-$gcc$-$ambi$} & 0.11 & 92.2 & 0.18 & 82.5 & 0.17 & 84.1 \\
\multicolumn{1}{l|}{$mbe$-$gcc$-$bin$}  & 0.12 & 91.6 & 0.20 & 79.8 & 0.24 & 77.2 \\ \hline
\multicolumn{1}{l|}{$mbe$-$ambi$} & \bf0.09 & 93.7 & \bf0.16 & \bf83.8 & \bf0.16 & \bf85.4 \\
\multicolumn{1}{l|}{$mbe$-$bin$}  & 0.10 & \bf93.8 & 0.18 & 81.8 & 0.22 & 78.5 \\
\multicolumn{1}{l|}{$mbe$-$mono$}  & 0.10 & 91.9 & 0.17 & 81.8 & 0.26 & 77.9 \\ 

\\
\multicolumn{1}{l|}{CRNN} & ER & F & ER & F & ER & F \\ \hline
\multicolumn{1}{l|}{$mbe$-$gcc$-$ambi$}  & 0.11 & 91.1 & 0.19 & 81.6 & 0.19 & 83.5 \\
\multicolumn{1}{l|}{$mbe$-$gcc$-$bin$}  & 0.12 & 92.3 & 0.21 & 78.8 & 0.26 & 79.0 \\ \hline
\multicolumn{1}{l|}{$mbe$-$ambi$}  & \bf0.10 & 92.8 & \bf0.18 & \bf82.5 & \bf0.17 & \bf83.7 \\
\multicolumn{1}{l|}{$mbe$-$bin$}  & 0.11 & \bf93.6 & 0.19 & 79.3 & 0.23 & 79.5 \\
\multicolumn{1}{l|}{$mbe$-$mono$}  & 0.12 & 91.9 & 0.18 & 80.6 & 0.28 & 78.3 \\
\end{tabular}
\end{table}

\begin{figure}[!b]
  \centering
  \centerline{\includegraphics[width=\linewidth]{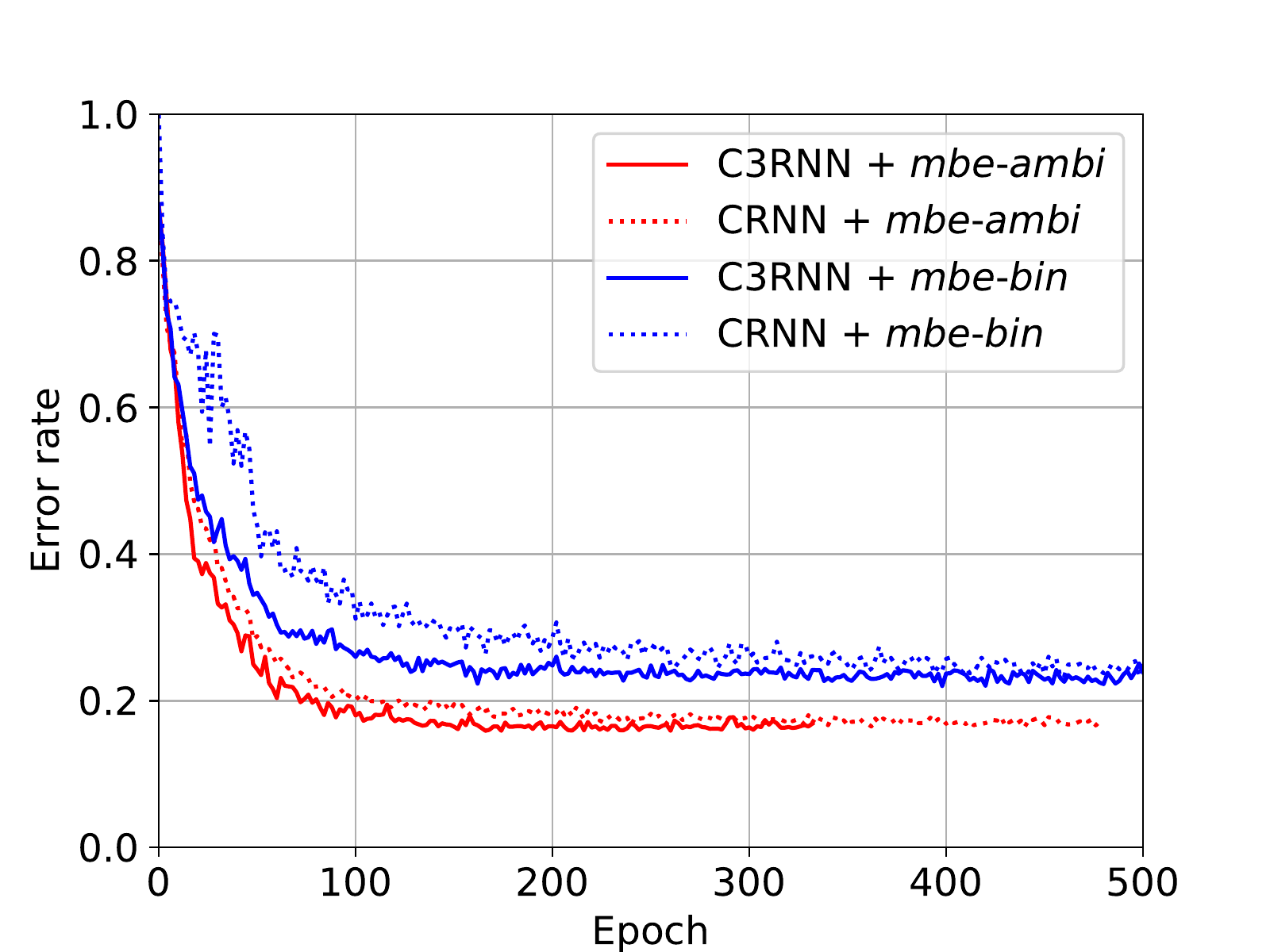}}
  \caption{The learning curve for the proposed C3RNN and baseline CRNN methods, for ambisonic ($mbe$-$ambi$) and binaural ($mbe$-$bin$) features of the synthetic $O6$ dataset. The proposed C3RNN achieves better error rate with a lower number of epochs, for both $mbe$-$ambi$ and $mbe$-$bin$ features.}
  \label{fig:loss-curve}
\end{figure}

The evaluation results for the proposed C3RNN using single-channel, binaural and Ambisonic audio $mbe$ and $gcc$ features for different polyphonic datasets are presented in Table~\ref{T:error}. Analyzing the performance of $mbe$ only features first, we see that for no polyphony ($O1$) the ER and F-scores are comparable for single ($mbe$-$mono$) and multichannel ($mbe$-$bin$ and $mbe$-$ambi$). With the increase in polyphony ($O3$ and $O6$), the ER and F scores of binaural and multichannel SED improves over the single-channel. Particularly, this improvement is significant for the dataset with highly overlapping sound events ($O6$). Concretely, using $mbe$-$ambi$ instead of $mbe$-$mono$ on $O6$ dataset gives 7.5\% improvement in F-score and 10\% in ER. A similar trend is observed using baseline CRNN for $mbe$ only feature and the results are comparable to proposed C3RNN. For the $mbe$-$mono$ feature the baseline CRNN and proposed C3RNN should ideally have the exact same scores across datasets since there is no additional inter-channel information for the C3RNN to learn from. The deviations seen in the metric scores are from the random initializations of the network even after averaging the scores from five separate runs on the cross-validation data. The actual improvement of using the proposed C3RNN over the baseline CRNN is achieved in training speed. As shown in Figure~\ref{fig:loss-curve}, C3RNN achieves better error rate with lower number of epochs for both $mbe$-$bin$ and $mbe$-$ambi$ features. The proposed C3RNN achieves this with exactly the same number of weights as the baseline CRNN, but with different convolution connections in CNN feature extraction layer.

Another observation from Table~\ref{T:error} for $mbe$-$mono$ feature and across methods is that the performance of SED drops with a higher number of overlapping sound events. Using multichannel features, especially $mbe$-$ambi$, the performance is comparable for up to three ($O3$) and six ($O6$) overlapping sound events datasets. In general, the $mbe$-$ambi$ is seen to perform better SED than the $mbe$-$bin$, which in turn performs better than $mbe$-$mono$. Additionally, the SED performance is seen to significantly improve with multichannel audio for sound scenes with highly overlapping sound events. This shows that using additional audio channel information definitely helps in more reliable and robust SED.

\begin{table}[!t]
\centering
\caption{Framewise accuracy (in \%) of recognizing the correct number of sound events in the synthetic $O6$ dataset.}
\label{T:conf_mat}
\begin{tabular}{lcccccccc}
            & \multicolumn{7}{c}{Number of overlapping sound events} \\ \cline{2-9}
\multicolumn{1}{l|}{C3RNN}       & 0      & 1      & 2     & 3     & 4     & 5     & 6 & \multicolumn{1}{|c}{Avg.}  \\ \hline
\multicolumn{1}{l|}{$gcc$-$ambi$} & 90.7 & 46.0 & 38.0 & 34.4 & 29.5 & 10.4 & 0.0 & \multicolumn{1}{|c}{35.6} \\\hline
\multicolumn{1}{l|}{$mbe$-$ambi$}   & \bf{92.7}   & \bf{66.9}   & \bf{56.3}  & \bf{47.7}  & \bf{34.7}  & \bf{16.3}  & 0.3 & \multicolumn{1}{|c}{\bf45.0}\\
\multicolumn{1}{l|}{$mbe$-$bin$} & 90.4   & 58.0     & 46.3  & 39.8  & 27.8  & 12.7  & \bf{0.6}  & \multicolumn{1}{|c}{39.4}\\
\multicolumn{1}{l|}{$mbe$-$mono$}         & 89.9   & 60.8   & 48.1  & 35.2  & 19.1  & 8.6   & 0.1 & \multicolumn{1}{|c}{37.4} \\ 

\\ 
CRNN        &        &        &       &       &       &       &      \\ \hline
\multicolumn{1}{l|}{$mbe$-$ambi$}  & \bf{93.5}   & \bf{66.4}   & \bf{56.5}  & \bf{47.3}  & \bf{32.4}  & \bf{15.7} & \bf{0.5}  & \multicolumn{1}{|c}{\bf44.6}\\
\multicolumn{1}{l|}{$mbe$-$bin$} & 92.8   & 60.6   & 47.7  & 42.9  & 29.1  & 12.3  & 0.2  & \multicolumn{1}{|c}{40.8}\\
\multicolumn{1}{l|}{$mbe$-$mono$}         & 90.8   & 59.6   & 49.9  & 34.1  & 18.4  & 9.7  & 0.4 & \multicolumn{1}{|c}{37.6}
\end{tabular}
\end{table}

Table~\ref{T:error} also reports the performance of using $mbe$ and $gcc$ features together ($mbe$-$gcc$). Since $gcc$ can only be extracted for more than one channel of audio, it reports results only for the binaural and Ambisonic versions of audio. In comparison to its respective $mbe$ only features, the evaluation metric scores are either comparable or worse for both C3RNN and baseline CRNN methods. To investigate this, and understand if using $gcc$ feature provides additional information to $mbe$, the experiment of estimating the number of sound sources per frame was carried out. An average accuracy of 35.6~\% was obtained in estimating the number of sound sources per frame using just $gcc$-$ambi$ (see Table~\ref{T:conf_mat}), while $mbe$-$ambi$ alone gave 45.0~\%. Similar results were obtained using binaural audio on the synthetic $O6$ dataset and the TUT-SED 2017 dataset (see Table~\ref{T:conf_mat_real}). 
% EDIT AP
%This shows that the usage of the $gcc$ feature is not providing any additional information over $mbe$ for the SED datasets studied in this paper. But previously, the $gcc$ feature has shown to be helpful for other binaural SED datasets~\cite{Adavanne_DCASE2016,Adavanne2017}, so this could just be a dataset- or audio-format specific result. In general, the inter-channel level difference of the $mbe$ feature which has been shown to capture the most information about the overlapping sound events for the binaural and Ambisonic format in this paper, would be insignificant for other formats such as a two spaced omnidirectional microphones, in which case the $gcc$ feature would be more helpful to provide useful information regarding the number of sources. 
Although the usage of $gcc$ feature in addition to $mbe$ has been shown to be helpful for other binaural SED datasets~\cite{Adavanne_DCASE2016,Adavanne2017}, the present results in Table~\ref{T:error} show that it does not provide any additional information for the SED datasets studied in this paper. The dominance of the $mbe$ features could be explained by the strong head shadowing effects at different source directions in binaural recordings and the spatial coincidence of Ambisonic recordings that encodes spatial information based only on inter-channel level differences. This dominance of $mbe$ feature may not hold for audio formats which rely on phase- or time-differences to encode directional information, with insignificant level differences. Audio captured with linear arrays or spaced omnidirectional microphones are examples of such audio formats and may benefit significantly from $gcc$ features instead of the level differences captured in $mbe$.

Among the audio features in Table~\ref{T:conf_mat}, we see that using multichannel features, especially $mbe$-$ambi$ significantly improves the accuracy of estimating overlapping sound events in comparison to single-channel $mbe$-$mono$ feature. In numbers, using $mbe$-$ambi$ instead of $mbe$-$mono$ in the proposed C3RNN method improves the performance of detection of three overlapping sources by 12.5~\% and four overlapping sources by 15.6~\%. This proves that using multichannel audio for SED helps recognize overlapping sound events better than single-channel audio.

The evaluation metric scores for the real-life recordings TUT-SED 2017 dataset is presented in Table~\ref{T:real_error}. The results are consistent with the results obtained with synthetic datasets. The performances of C3RNN and CRNN are comparable, and the multichannel feature $mbe$-$bin$ achieves better SED than single-channel $mbe$-$mono$ feature. Additionally, the error rate obtained using the proposed C3RNN and $mbe$-$bin$ feature beats the current top score of 0.50~\cite{Adavanne_DCASE2017_binaural} on this benchmarking dataset.

\begin{table}[!t]
\centering
\caption{Framewise accuracy (in \%) of recognizing the correct number of sound events in the TUT-SED 2017 dataset.}
\label{T:conf_mat_real}
\begin{tabular}{l|cccc|c}
 & \multicolumn{5}{c}{Number of overlapping sources} \\ \cline{2-6}
C3RNN & 0 & 1 & 2 & 3 & Avg. \\ \hline
$mbe$-$bin$ & \bf70.1 & \bf70.2 & \bf73.1 & \bf16.4 & \bf57.5 \\
$gcc$-$bin$ & 62.2 & 64.6 & 39.4 & 2.0 & 42.1
\end{tabular}
\end{table}

\begin{table}[!t]
\centering
\caption{The evaluation metric scores for the sound event detection task using the proposed C3RNN and baseline CRNN for the TUT-SED 2017 dataset.}
\label{T:real_error}
\begin{tabular}{l|cc|cc}
 & \multicolumn{2}{c}{C3RNN} & \multicolumn{2}{c}{CRNN} \\\cline{2-5}
 & ER & F & ER & F \\ \hline
$mbe$-$bin$ & \bf0.35 & \bf67.5 & \bf0.37 & \bf64.8 \\
$mbe$-$mono$ & 0.38 & 64.1 & 0.39 & 63.3 \\
\end{tabular}
\end{table}

\section{Conclusion} 
\label{sec:conclusion}
In this paper, we proposed a stacked convolutional and recurrent neural network with inter- and intra-channel convolutions in the first layer (C3RNN) for the multichannel sound event detection (SED) task. The inter- and intra-channel convolutions were implemented using a 3D convolutional neural network (CNN) layer. It was shown that the proposed C3RNN method learns to recognize overlapping sound events from multichannel features faster than the state of the art baseline multichannel SED method with exactly the same number of parameters, and further performs better SED with fewer number of training epochs. In the multichannel SED task, for the SED datasets used in the paper, it was shown that the generalized cross-correlation with phase transform feature was not providing any additional information to the standard multichannel log mel-band energy feature.

Additionally, we proposed to assess the performance of using multichannel audio for polyphonic SED. The study was carried out on four datasets- a) one real-life recording TUT-SED 2017 development dataset, and three synthetic datasets with b) no temporally overlapping, c) up to three temporally overlapping and d) up to six temporally overlapping sound events. Each of the recordings in the synthetic datasets was in three formats, four-channel first-order Ambisonics, binaural and single-channel, whereas the real-life dataset had just the binaural and single-channel audio versions. We performed SED individually on these datasets using the proposed C3RNN. In comparison to using a single-channel, we observed that by using multichannel audio, the overall F-score improved by 7.5~\%, overall ER improved by 10~\% and 15.6~\% more sound events were recognized in time frames with four overlapping sound events. In conclusion, multichannel audio definitely improves the SED performance over using single-channel audio; and the collection of such a real-life multichannel audio dataset is worth the effort.

\bibliographystyle{IEEEtran}
\bibliography{refs}
\end{document}